\documentclass[9pt,twocolumn,superscriptaddress]{revtex4}

\usepackage{multirow}
\usepackage{xcolor}
\usepackage{bm,amsmath}
\usepackage{amssymb}
\usepackage{graphicx}
\usepackage{array}
\definecolor{darkblue}{rgb}{0,0,0.6}
\usepackage[colorlinks,linkcolor=darkblue,citecolor=darkblue,urlcolor=darkblue]{hyperref}
\usepackage{pdfpages}
\renewcommand{\r}{\textcolor{black}}

\newcommand{\Teff}{T_\text{eff}}

\begin{document}

\title{Scaling the glassy dynamics of active particles: Tunable fragility and re-entrance}

\author{Puneet Pareek}
\affiliation{Tata Institute of Fundamental Research, Hyderabad - 500046, India}

\author{Peter Sollich}
\affiliation{Institute for Theoretical Physics, University of G{\"{o}}ttingen, Friedrich-Hund-Platz 1, 37077 G{\"{o}}ttingen, Germany}

\author{Saroj Kumar Nandi}
\email{saroj@tifrh.res.in}
\affiliation{Tata Institute of Fundamental Research, Hyderabad - 500046, India}

\author{Ludovic Berthier}
\affiliation{Gulliver, CNRS UMR 7083, ESPCI Paris, PSL Research University, 75005 Paris, France}


\begin{abstract}
Understanding the influence of activity on dense amorphous assemblies is crucial for biological processes such as wound healing, embryogenesis, or cancer progression. Here, we study the effect of self-propulsion forces of amplitude $f_0$ and persistence time $\tau_p$ in dense assemblies of soft repulsive particles by simulating a model particle system that interpolates between particulate active matter and biological tissues. We identify the fluid and glass phases of the three-dimensional phase diagram obtained by varying $f_0$, $\tau_p$, and the packing fraction $\phi$. The morphology of the phase diagram directly accounts for a non-monotonic evolution of the relaxation time with $\tau_p$, which is a direct consequence of the crossover in the dominant relaxation mechanism, from glassy to jamming. A second major consequence is the evolution of the glassy dynamics from sub-Arrhenius to super-Arrhenius. We show that this tunable glass fragility extends to active systems analogous observations reported for passive particles. This analogy allows us to apply a dynamic scaling analysis proposed for the passive case, in order to account for our results for active systems. Finally, we discuss similarities and differences between our results and recent findings in the context of computational models of biological tissues.
\end{abstract}

\maketitle

\section{Introduction}

The effects of active processes on glassy dynamics have fundamental importance in several biological processes, such as wound healing~\cite{poujade2007,vishwakarma2020}, embryogenesis~\cite{mongera2018,park2016,atia2021,Tambe2011,schotz2013}, asthma development~\cite{park2015}, or cancer progression~\cite{friedl2003a,friedl2009,Malmi-Kakkada2018,Sinha2020,palamidessi2019,streitberger2020,mitchel2020}. 
The collective cellular dynamics during these processes exhibit a transition from a solid-like to a fluid-like state, although the static properties remain nearly the same~\cite{Angelini2011,Berthier2011,atia2021,activereview,Berthier2019,janssen2019Review}. The characteristics of the resulting glassy dynamics are broadly similar to those in equilibrium particulate systems~\cite{Berthier2011,Berthier2019,activereview,higler2018}, although novel features may arise from the non-equilibrium nature of the systems. Indeed, cells are active and cellular systems are constantly evolving far from equilibrium~\cite{sriramreview,sriramrmp}. Cells can also change their characteristics during various processes~\cite{thiery2002,thiery2009}, and they are extended objects for which cellular shapes closely correlate with biological functions~\cite{park2015,sailem2017,atia2018,poujade2007,chen1997}.
Geometric features can also differ from passive systems for the same reason; epithelial systems, for example, are always confluent, so that the cells entirely cover the space and the packing fraction may not be the most relevant parameter~\cite{atia2021}. 

The complexity of active biological systems makes it challenging to gain insights into the key mechanisms driving their glassy dynamics. 
Simulation studies of model cellular systems have been instrumental here, showing a rigidity transition akin to the jamming transition~\cite{bi2015,bi2016,park2015}. Furthermore, several computational studies have demonstrated that these systems exhibit nontrivial glassy behavior, such as sub-Arrhenius relaxation dynamics~\cite{sussman2018,li2021,sadhukhan2021}, possibly crossing over to super-Arrhenius behavior as the shape index (which controls cell shape) varies~\cite{sadhukhan2021,li2021,sadhukhan2024}. It would be useful to elucidate whether these features are specific to biological tissues and vertex models, or can be observed in simpler models of active particles.

Active systems composed of self-propelled particles cannot capture all microscopic details of biological tissues, but may nevertheless provide useful insights into the competition between crowding and activity. Beyond this they can also reveal what features are specific to confluent models with many-body interactions, compared to particle models with pairwise forces. In the latter models, particles self-propel under the influence of self-propulsion forces of amplitude $f_0$ and persistence time $\tau_p$~\cite{sriramreview,sriramrmp,activereview,Berthier2019,janssen2019Review,Mandal2020,mandal2020aging,wiese2023,szamel2024}. In fact, several biological systems can be conveniently modelled as dense systems of self-propelled particles on different lengthscales, such as for example collections of cells~\cite{Angelini2011,Garcia2015}, the cytoplasm~\cite{zhou2009,parry2014,nishizawa2017}, aggregates of organisms~\cite{gravish2013,lama2024}, and synthetic active matter~\cite{deseigne2010,Arora2022,Palacci2013,lam2015}. Several experiments~\cite{fabry2001,Angelini2011,lenorman2007,Arora2022} and simulation studies~\cite{Berthier2014,Mandal2016,mandal2017,Flenner2016,Berthier2017} have shown that active systems in their dense regime exhibit many characteristics similar to glasses~\cite{Berthier2019,janssen2019Review,atia2021,activereview}, such as the anomalously slow two-step relaxation dynamics \cite{park2015} and dynamical heterogeneity \cite{dhbook}. Recent numerical studies have also revealed nontrivial effects of activity on the glassy dynamics, such as a shifted glass transition point~\cite{Berthier2013,Berthier2014,Nandi2017}, changing fragility~\cite{Mandal2016,Flenner2016,Nandi2018,dey2024}, and re-entrant relaxation dynamics~\cite{Berthier2017,Flenner2016,Debets2021}. These active particle systems thus provide a rich yet still relatively simple framework for understanding the role of nonequilibrium fluctuations on the glassy dynamics.

There are several variants of active matter models in the literature~\cite{Marchetti2013}, sometimes categorized into dry and wet active matter. In the former, the surrounding fluid is either absent or treated implicitly within the models. In the latter, the surrounding fluid is explicitly taken into account. Since glassy effects arise in the long-time dynamics, we expect the hydrodynamic role of the surrounding fluid to be insignificant, as it affects the physics on much shorter timescales. We thus believe that the dry-wet distinction is less relevant in the present context~\cite{Klongvessa2019,Klongvessa2022}. Similarly, dimensionality effects are not expected to play an important role for the specific questions addressed in our study.

Theories of the equilibrium glass transition, such as the mode-coupling theory~\cite{Berthier2013,Szamel2015,Szamel2016,Feng2017,Liluashvili2017,Nandi2017,paul2023,debets2021_mct,paul2023,debets2023,kolya2024,pandey2023} and the random first-order transition approach~\cite{Nandi2018,Mandal2022,sadhukhan2021,sadhukhan2024} have been extended to active glasses to rationalize numerical and experimental results. In fact, several aspects of activity-driven glassy dynamics remain equilibrium-like at a suitably defined effective temperature \cite{Mandal2016,Nandi2018}. However, there are also a number of nontrivial aspects that have been reported, such as re-entrant glassy dynamics \cite{Berthier2017,Debets2021,Arora2022}, change of glass fragility \cite{Mandal2016,Flenner2016,Nandi2018,dey2024}, and emerging velocity correlations \cite{Szamel2015}. Furthermore, the specific role of $\tau_p$ on the glassy dynamics has not been elucidated fully. Crucially, how to extend and connect these results to studies of more complex biological systems remains unclear.

To address all these issues, we investigate active glassy dynamics via computer simulations of a system of soft repulsive particles. This model smoothly connects and unifies distinct physics in appropriate limits. In the limit of zero activity, the finite-temperature ($T$) glassy behavior is governed by a ``glass point'' in the limit of small but nonzero $T$, and a jamming transition at $T=0$ itself. One can thus explore both jamming and glassy physics within the same system~\cite{Berthier2009,Berthier2009-pre,Berthier2010,jacquin2011microscopic,Adhikari2023}. On the other hand, the same system in the very dense limit is a good approximation of confluent cellular systems. Between these two limits the system describes the active matter physics of particulate systems, where qualitatively different regimes can be expected as the persistence time is varied. Therefore, this is a fairly rich model, which as we will demonstrate exhibits several non-trivial features that have also been reported in biological tissues. Our goal is, then, to provide a physical interpretation of the observed behaviors, in the context of a particle model that is relatively easy to study and understand. This approach will also be useful to guide and interpret future studies of biological materials and tissue models, for instance to disentangle specific effects due to the confluent nature of tissues or the possible influence of many-body forces.

Our manuscript is organised as follows.
In Sec.~\ref{model}, we introduce the model and its control parameters and explain the various limiting situations it describes. 
In Sec.~\ref{characterising}, we present the broad glassy features for a selected set of parameters. 
In Sec.~\ref{phasedia}, we collect our results to contruct the complete three-dimensional phase diagram of the model delimiting fluid and glass phases. 
In Sec.~\ref{reentrance}, we show that the location of the glass transition has a non-trivial dependence on the persistence time at fixed driving amplitude, leading to re-entrant glassy dynamics.   
In Sec.~\ref{fragility}, we show that the system displays a crossover from sub-Arrhenius to super-Arrhenius dynamics as density and persistence time are increased. 
In Sec.~\ref{BerthierWitten}, we extend a dynamic scaling analysis proposed for equilibrium soft spheres to our soft persistent particles. 
In Sec.~\ref{discussion}, finally, we discuss our results and provide some perspectives regarding biological tissues. 

\section{Model for soft active particles}

\label{model}

We consider a three-dimensional 50:50 binary mixture of particles of two types A and B, interacting via the Weeks-Chandler-Andersen purely repulsive potential. We expect that similar results will hold in two dimensions, as the effects of dimensionality on glassy dynamics are well understood~\cite{flenner2015fundamental}. We assume that thermal fluctuations are not relevant and set the Brownian temperature to $T=0$. Instead, the particles are driven by non-thermal self-propulsion forces, for which we use the AOUP (active Ornstein-Uhlenbeck particle) activity model (see supplementary material (SM), Sec.~S1 for further details). Previous studies have demonstrated that similar physics would be obtained independently of the specific model chosen for the self-propulsion or the repulsive pairwise interaction~\cite{Debets2022}.

We use molecular dynamics simulations to evolve the particle position $\mathbf{r}_i$ of the $i$th particle using the following equation of motion:
\begin{equation} \label{particle_pos}
	\dot{\mathbf{r}}_i=\xi_0^{-1}\left[\mathbf{F}_i+\mathbf{f}_i \right],
\end{equation}
where $\xi_0$ the friction coefficient, $\mathbf{F}_i$ is the interaction force felt by the $i$th particle from the other particles, and $\mathbf{f}_i$ is the active self-propulsion force. The latter follows an Ornstein-Uhlenbeck stochastic process
\begin{equation}\label{active_force}
	\tau_p \dot{\mathbf{f}}_i=-\mathbf{f}_i+\boldsymbol{\eta}_i,
\end{equation}
with $\langle \boldsymbol{\eta}_i(t) \boldsymbol{\eta}^{\rm T}_j(t')\rangle=2 f_0^2 \mathbf{I}\delta_{ij}\delta(t-t')$, where ``T'' denotes the transpose and $\mathbf{I}$ the identity matrix. We present results using $\xi_0 \sigma^2_{BB}/\epsilon$ as the time unit, with $\sigma_{BB}$ the diameter of B-type particles that sets the unit length, and $\epsilon$ the energy scale of the pair potential.

The self-propulsion force $\mathbf{f}_i$ in Eq.~(\ref{particle_pos}) is stochastic with zero mean and correlator $\langle \mathbf{f}_i(t) \mathbf{f}^{\rm T}_j(0)\rangle =(f_0^2/\tau_p)\exp[-t/\tau_p] \mathbf{I} \delta_{ij}$. It thus has a typical amplitude $ |\mathbf{f}_i | \sim f_0 / \sqrt{\tau_p}$ and fluctuations that are correlated over a time of order $\tau_p$. In the limit of very small persistent times, $\tau_p \to 0$, the propulsion forces thus become equivalent to a thermal white noise, with $f_0^2$ playing the role of a temperature. In the opposite limit of large persistence times, the propulsions become nearly constant random driving forces of amplitude $f_0 / \sqrt{\tau_p}$.

\begin{figure*}
	\includegraphics[width=18cm]{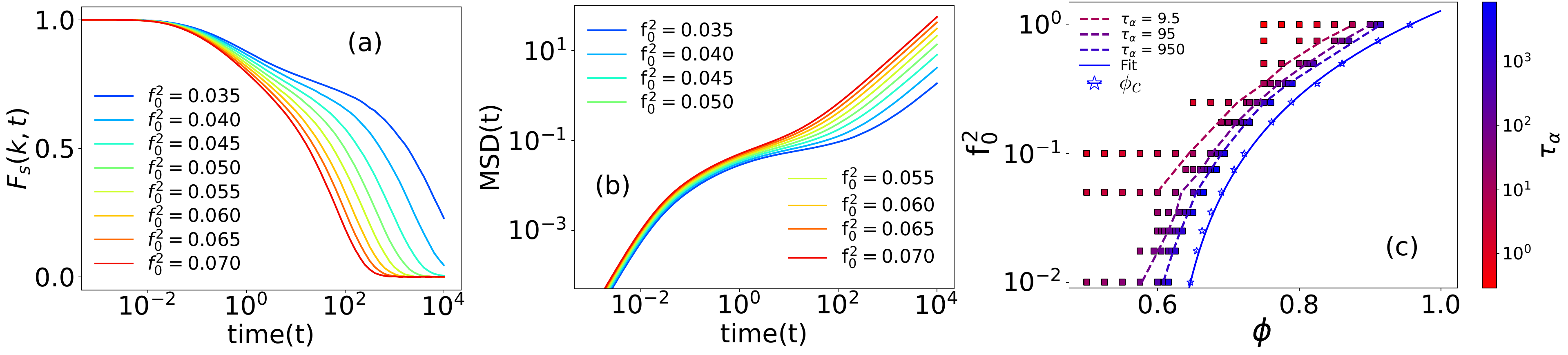}
	\caption{Characterization of the active glassy dynamics. (a) The two-step decay of $F_s(k,t)$ becomes faster with increasing $f_0$ at constant $\phi=0.65$ and $\tau_p=10^{-2}$. (b) The corresponding MSD has sub-diffusive to diffusive crossover at intermediate times, and particle motion is faster with increasing $f_0$. (c) The liquid-glass $(f_0,\phi)$ phase diagram for $\tau_p=10^{-2}$. Squares represent the simulated state points with the corresponding $\tau_\alpha$ color-coded, and dashed lines representing iso-$\tau_\alpha$ contours. Stars are the $\phi_c$ values obtained using Eq.~(\ref{eq:VFT}), while the solid line connecting them is a fit to Eq.~(\ref{f0eq}). } 
	\label{relaxation_dynamics}
\end{figure*}

To smoothly interpolate between these two limits, it is useful to introduce an effective temperature, $\Teff$~\cite{Nandi2017,Nandi2018,cugliandolo2011,cugliandolo2019,petrelli2020,nandi2018Teff}, with
\begin{equation}
  \Teff = \frac{f_0^2}{1+G\tau_p},
\label{eq:Teff}
\end{equation}
where $G$ is a constant~\cite{Nandi2017}. This effective temperature reduces to $\Teff \sim f_0^2$ for small persistence times, and to $\Teff \sim f^2_0/ \tau_p$ for larger $\tau_p$. The expression in Eq.~(\ref{eq:Teff}) can be analytically justified by considering the position fluctuations of a single AOUP in a harmonic potential~\cite{szamel2014self}.

The concept of an effective temperature has a long history in both active and glass matter~\cite{cugliandolo2011}. For active glasses, an effective temperature naturally emerges in the long-time dynamics~\cite{Berthier2013,Levis2015,Nandi2017}. It has even been shown that the relaxation dynamics for the active system agrees well with the mode-coupling scaling description using $\Teff$. Using this effective temperature description for the relaxation dynamics and comparing it with the mode-coupling theory scaling relations (see SM Fig.~S5), we estimate $G\simeq 0.5$ for the current system. For the present purposes, $\Teff$ is a useful quantity as it reduces to a genuine thermodynamic temperature in the small $\tau_p$ limit, but our results do not rely on any thermodynamic interpretation far from equilibrium.

The three control parameters of the model are the persistence time $\tau_p$, the amplitude $f_0$ in the noise term in Eq.~(\ref{active_force}) and the volume fraction $\phi$. We will systematically vary them to fully characterise the part of the phase diagram where the system becomes glassy. It is first interesting to consider the various physical limits captured by this model.

{\it Small persistence time, $\tau_p \to 0$.} The phase diagram becomes effectively two-dimensional in the equilibrium limit $\tau_p \to 0$, where the system becomes equivalent to a Brownian system with the two control parameters temperature and density, ($\Teff \sim f_0^2$, $\phi$), as usual in equilibrium fluids. The smooth approach to the Brownian limit makes the AOUP model in Eqs.~(\ref{particle_pos}, \ref{active_force}) appealing. In this limit, the physics becomes strictly equivalent to the one of soft repulsive spheres at thermal equilibrium, as studied extensively in Ref.~\cite{Berthier2009-pre}. 

{\it Small force amplitude, $f_0 \to 0$.} For finite $\tau_p$, another interesting limit is the one where $f_0 \to 0$. In this limit, the amplitude of the self-propulsion force $\mathbf{f}_i$ in Eq.~(\ref{particle_pos}) becomes much smaller than that of the  interaction forces $\mathbf{F}_i$, so that particles will no longer interpenetrate (unless forced to do so by a high volume fraction $\phi$). As a consequence, the system behaves as self-propelled hard spheres with a finite persistence time $\tau_p$. The phase diagram is again two-dimensional, with $(\tau_p, \phi)$ as control parameters. Such self-propelled hard spheres have been
studied in several past works~\cite{ni2013pushing,Berthier2014} and
undergo a non-equilibrium glass transition, at a volume fraction $\phi_d(\tau_p)$ that depends continuously on the persistence time. These special ``glass points'' in the phase diagram will play a prominent role in our dynamic scaling analysis in Sec.~\ref{BerthierWitten}.

{\it Large persistence time, $\tau_p \to \infty$.} The final interesting limit is that of infinitely persistent particles obtained when $\tau_p \to \infty$. In that limit, particles are driven by frozen random forces whose amplitudes $|f_i| \sim f_0 / \sqrt{\tau_p}$ become vanishingly small compared to interparticle forces, in the absence of any other source of fluctuations. By construction, the system undergoes in this limit a jamming transition at a packing fraction $\phi_J$, between a flowing fluid and a solid phase. We expect the dynamics in this ``jamming'' limit to differ qualitatively from the glassy dynamics observed near glass transitions~\cite{ikeda2012}.   

\section{Characterising the glassy dynamics}

\label{characterising}

We characterize the glassy dynamics via the self-intermediate scattering function, $F_s(k,t)$, at wavevector $k$ and time $t$, defined as
\begin{equation}
  \label{Fskt}
	F_{s}(k,t) = \frac{1}{N}\Big \langle \sum_{i = 1}^{N} e^{\dot{\iota} \mathbf{k}\cdot (\mathbf{r}_{i}(t+t_0) - \mathbf{r}_{i}(t_0))} \Big \rangle,
\end{equation}
and by the mean-squared displacement
\begin{equation}\label{MSDoft}
	\text{MSD}(t)=\Big \langle\frac{1}{N} \sum_{i=1}^N [\mathbf{r}_i(t+t_0)-\mathbf{r}_i(t_0)]^2\Big \rangle,
\end{equation}
where $N$ is the total number of particles, $\mathbf{r}_i(t)$ is the position of the $i$th particle at time $t$, and $\langle\ldots\rangle$ denotes an average over independent configurations and over the time origins $t_0$ taken at steady state. For convenience we choose $k$ as the position of the first peak of the static structure factor of the passive system at $\phi=0.65$; this first peak position does not change significantly at other values of $\phi$ [SM Fig.~S1]. We first ask how the dynamics varies with $f_0$ at constant $\tau_p$, and then explore the evolution with $\tau_p$.

We show the time dependence of $F_s(k,t)$ and MSD($t$) at constant $\phi=0.65$ and $\tau_p=10^{-2}$ with varying $f_0^2$ in Figs.~\ref{relaxation_dynamics}(a, b). The self-intermediate scattering function $F_s(k,t)$ first decays towards a non-zero plateau value and at later times to zero, see Fig.~\ref{relaxation_dynamics}(a). The plateau region of $F_s(k,t)$ corresponds to sub-diffusive motion in the MSD, which only becomes diffusive at much longer times, see Fig.~\ref{relaxation_dynamics}(b). As $f_0$ increases, the decay of $F_s(k,t)$ and the transition in the MSD from sub-diffusive to diffusive behavior occur at shorter times. Overall the dynamical characteristics of $F_s(k,t)$ and MSD with increasing $f_0$ are similar to those of equilibrium supercooled fluids with increasing temperature $T$. For further quantification one can define the relaxation time $\tau_\alpha$ as the value of $t$ for which $F_s(k,t)$ decays to $1/e$, explicitly $F_s(k,\tau_\alpha)=1/e$. From Fig.~\ref{relaxation_dynamics}(a), it is then clear that $\tau_\alpha$ decreases as $f_0$ increases. We have explored the dynamics at several other packing fractions $\phi$, and the qualitative behavior remains the same.

\begin{figure*}
	\centering
	\includegraphics[width=14.6cm]{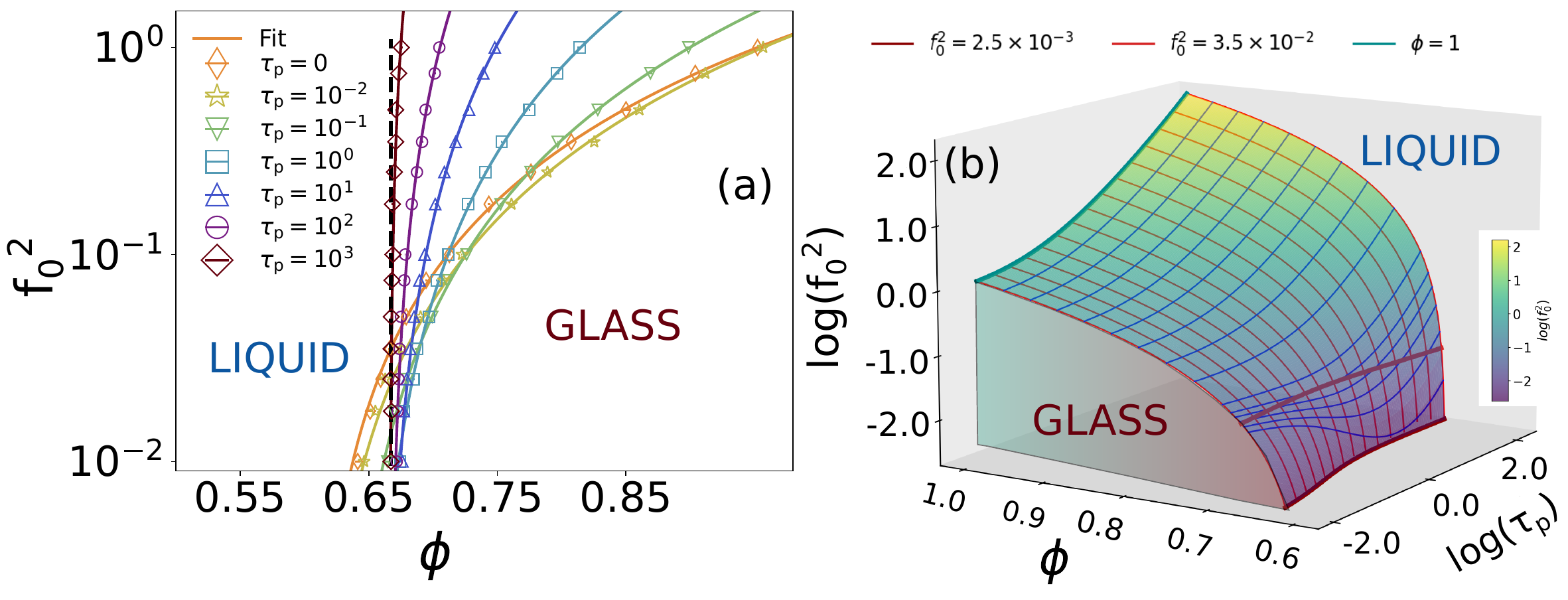}
	\caption{Constructing the three-dimensional phase diagram.
		(a) The liquid-glass critical lines (symbols) determined via Eq.~(\ref{eq:VFT}) for different values of $\tau_p$. These lines are themselves fitted to a power law form, Eq.~(\ref{f0eq}), shown as lines. Vertical dashed line indicates the limit $\phi_d(\tau_p \to \infty)$. 
		(b) The three-dimensional liquid-glass phase diagram can then be reconstructed from the fitted analytical expressions, with the glass phase occurring for large $\phi$ and low $f_0$, with a non-trivial evolution with $\tau_p$. Two iso-$f_0^2$ lines are shown; the one corresponding to $f_0^2=3.5\times 10^{-2}$ is non-monotonic.}
	\label{threedphasedia}
\end{figure*}

We now characterize the liquid-glass phase boundary at constant $\tau_p$. For this purpose, we first obtain $\tau_\alpha$ for several values of $\phi$ at constant $f_0$. Figure~\ref{relaxation_dynamics}(c) shows the state points we have explored and the corresponding values of $\tau_\alpha$. We then use a logarithmic fit,
\begin{equation}
  \ln \tau_\alpha=A+B/(\phi_c-\phi),
\label{eq:VFT}
\end{equation}
using $A$, $B$, and $\phi_c$ as fit parameters, to obtain the liquid-glass critical point $\phi_c(f_0^2,\tau_p)$ for fixed $f_0$ and $\tau_p$.
The expression~(\ref{eq:VFT}) is formally analogous to the Vogel-Fulcher-Tammann (VFT) relation describing the relaxation time of molecular liquids, and implies that the system becomes a glass at the point where $\tau_\alpha$ diverges, i.e.\ for $\phi=\phi_c(f_0^2,\tau_p)$. We repeat the above fitting procedure for a range of $f_0$ and use the results to construct the liquid-glass critical line in the $(f_0, \phi)$ plane at fixed $\tau_p$, as shown in Fig.~\ref{relaxation_dynamics}(c) for the specific value $\tau_p = 10^{-2}$. For this small $\tau_p$, the critical glass line in the $(f_0, \phi)$ plane is by construction very close to the equilibrium glass line obtained for $\tau_p \to 0$, with $f_0^2$ playing the role of temperature~\cite{Berthier2009,Berthier2009-pre}.

The results in this section are broadly consistent with existing literature on active glassy systems, where the effects of activity remain equilibrium-like at a suitably defined effective temperature as long as $\tau_p$ is small~\cite{Mandal2016,Flenner2016,Nandi2017,Nandi2018}. In the next sections, we explore a much broader range of control parameters, which will allow us to reveal additional effects that are non-trivial.

\section{Three-dimensional phase diagram}

\label{phasedia}

We now construct the three-dimensional liquid-glass phase diagram in the full control parameter space $(f_0, \tau_p, \phi)$ to understand how the persistence time $\tau_p$ changes the results set out in Sec.~\ref{characterising} above. For this purpose, we obtain the critical lines $\phi_c(f_0^2,\tau_p)$ by repeating the fitting procedure shown in Fig.~\ref{relaxation_dynamics}(c) for several values of $\tau_p$. This numerical exploration of the three-dimensional space represents a significant computational effort, given that it requires simulations scanning a range of volume fractions $\phi$ for each pair of values $(f_0,\tau_p)$. This large computational effort is useful, as it allows us to recover within a single approach several results obtained independently in various limits, which we can rationalise and organise in the scaling approach developed below. The symbols in Fig.~\ref{threedphasedia}(a) show the resulting critical lines, $\phi_c(f_0^2,\tau_p)$ in the $(f_0, \phi)$ plane for several values of $\tau_p$. Similar data were obtained in Ref.~\cite{Berthier2017} using the diffusivity, but the detailed characteristics and consequences were not fully explored.

Several features are immediately apparent from the data in Fig.~\ref{threedphasedia}(a). First, $\phi_c(f_0^2,\tau_p)$ displays a non-trivial evolution with $\tau_p$. As $\tau_p$ increases, the critical lines become more vertical, which implies that the glass transition line becomes less sensitive to $\phi$ for larger persistence times. Also, the $\phi_c(f_0^2 \to 0,\tau_p)$ values shift to higher packing fractions and appear to saturate at very large $\tau_p$. Finally, the evolution of $\phi_c$ with $\tau_p$ exhibits opposite trends for small and large values of $f_0$, respectively. We will relate these observations below to the physical behaviour of the system.

To proceed, we first propose an analytical description of the critical surface delimiting fluid and glass phases. It is convenient to first introduce the notation
\begin{equation}
\phi_d(\tau_p) \equiv \phi_c(f_0^2 \to 0,\tau_p), 
\end{equation}
which defines the location of the glass transition in the self-propelled hard sphere limit. We then fit the critical lines in Fig.~\ref{threedphasedia}(a) with a power law 
\begin{equation}
  \label{f0eq}
	f_0^2 = a [\phi_c(f_0^2,\tau_p)-\phi_{d}]^{2/\beta},
\end{equation}
where the prefactor $a$, the exponent $\beta$ and the critical density $\phi_d$ are used as fit parameters for each $\tau_p$ value. The resulting fitted functions are displayed as lines in Fig.~\ref{threedphasedia}(a). The fits confirm that $\phi_d(\tau_p)$ first increases with $\tau_p$ and then saturates at larger $\tau_p$ (see SM Fig.~S4). The prefactor $a(\tau_p)$ monotonically increases with increasing $\tau_p$ (see SM Fig.~S3). On the other hand, the exponent $\beta(\tau_p)$ has a weak non-monotonic $\tau_p$-dependence (see SM Fig.~S3). In the SM, we propose empirical fitting forms to represent the $\tau_p$-dependence of these three constants, which work well (see SM, Sec.~S4).

We are finally in a position to collect the above fits of our simulation data into a liquid-glass surface in the three-dimensional phase diagram shown in Fig.~\ref{threedphasedia}(b), using as axes $(f_0^2, \phi, \tau_p)$. The glass phase is located in the large-$\phi$, low-$f_0$ corner of this parameter space, with non-trivial evolution with $\tau_p$ and opposite trends at large and small $\tau_p$ values. In the next two sections, we explore two major consequences of the shape of this phase diagram.

\section{Re-entrant relaxation dynamics}

\label{reentrance}

\begin{figure*}
	\centering
	\includegraphics[width=17cm]{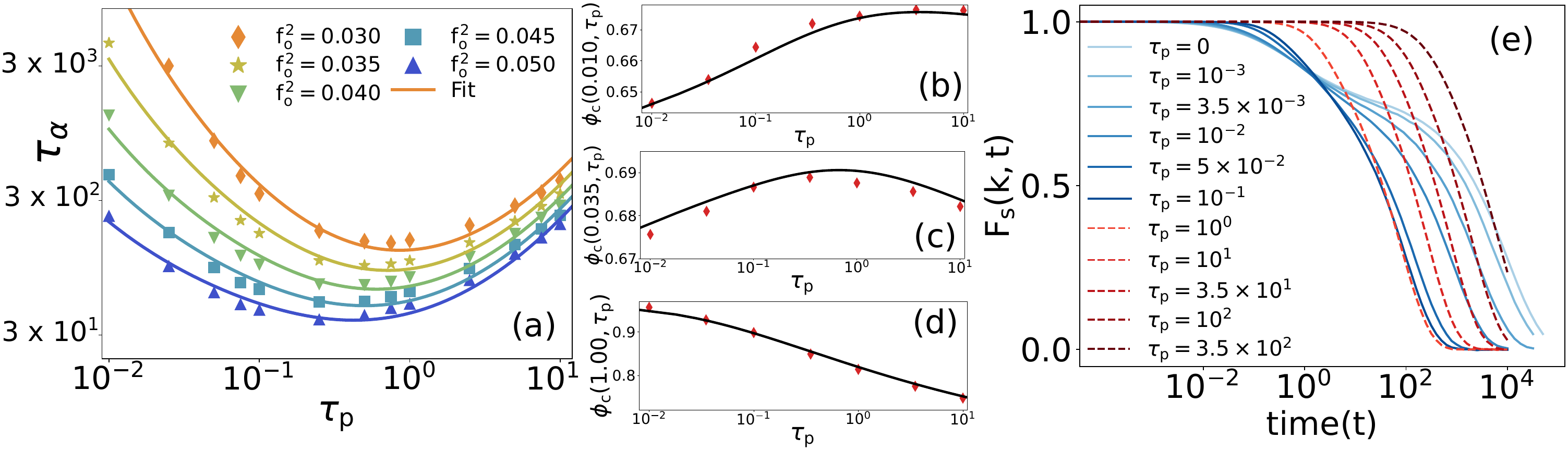}
	\caption{Re-entrant glassy dynamics.
		(a) Non-monotonic behaviour of $\tau_\alpha$ as a function of $\tau_p$ at various $f_0^2$ and $\phi=0.65$. The lines combine Eqs.~(\ref{eq:VFT}, \ref{f0eq}) (see SM Table S1).
		(b, c, d) The dependence of $\phi_c(f_0^2,\tau_p)$ on $\tau_p$ for a small, an intermediate and a large value of $f_0^2$, showing a change from a monotonically increasing to a monotonically decreasing trend, with non-monotonic variation for the intermediate $f_0^2$. Lines represent the analytical description in Eq.~(\ref{critsurf}).
		(e) Time decay of $F_s(k,t)$ for $\phi=0.65$ and $f_0^2=0.045$ and various $\tau_p$. The intermediate time plateau at small $\tau_p$, i.e.\ in the glassy regime, disappears at large $\tau_p$ in the jamming regime. }
	\label{reentranceVFT}
\end{figure*}

We first establish the presence of re-entrant glassy dynamics, where $\tau_\alpha$ has a non-monotonic variation with changing $\tau_p$, as a consequence of the $\tau_p$ dependence of the critical surface constructed in Fig.~\ref{threedphasedia}(b). We present in Fig.~\ref{reentranceVFT}(a) the evolution of $\tau_\alpha$ as a function of the persistence time $\tau_p$ for a range of values of $f_0^2$. These data indeed show that $\tau_\alpha$ has a non-trivial non-monotonic evolution with the persistence time $\tau_p$ at fixed values of $\phi$ and $f_0^2$.

Qualitatively this behavior arises because of the opposite trends mentioned above: the critical liquid-glass density increases with $\tau_p$ at small $f_0$, whereas it decreases at large $f_0$. These contrasting trends create an intermediate range of $f_0$ where the critical density is non-monotonic in $\tau_p$. We show this explicitly in Figs.~\ref{reentranceVFT}(b, c, d) where we follow the evolution with $\tau_p$ of $\phi_c(f_0^2,\tau_p)$ for a small, intermediate and large value of $f_0$ and observe increasing, non-monotonic, and decreasing variation with $\tau_p$, respectively. 

To describe this re-entrant behavior more quantitatively, we start from Eq.~(\ref{f0eq}) and rewrite the equation describing the critical glass surface $\phi_c(f_0^2,\tau_p)$ as
\begin{align}\label{critsurf}
	\phi_c(f_0^2,\tau_p)=[f_0^2/a]^{\beta/2} + \phi_d(\tau_p).
\end{align}
This critical density is, by construction, controlled by the second term when $f_0 \to 0$, so that $\phi_c$ increases with $\tau_p$ along with $\phi_d$. For large $f_0$, on the other hand, the first term dominates and $\phi_c$ then decreases with $\tau_p$ due to the increase of $a$ (recall Fig.~S3 in SM). The lines in Figs.~\ref{reentranceVFT}(b, c, d) confirm that the empirical description of the glass surface obtained in Sec.~\ref{phasedia} describes the numerical data, and thus the re-entrant dynamics, very well. 

To conclusively establish that the shape of the critical surface does describe the re-entrant behavior, we use Eq.~(\ref{eq:VFT}). This expression can be obtained within the random-first order transition theory, which constitutes a microscopic theory of the glass transition~\cite{lubchenko2007theory,Kirkpatrick2015} that has been extended for active systems~\cite{Nandi2018,Mandal2022}. Using $\phi_c(f_0^2,\tau_p)$ in Eq. (\ref{eq:VFT}), we obtain $\tau_\alpha$ as
	\begin{equation}\label{reentrant}
		\ln\tau_\alpha = A + \frac{B}{c[f_0^2/a(\tau_p)]^{\beta(\tau_p)/2} + \phi_d(\tau_p) - \phi},
	\end{equation}
where we have included the fitting parameter $c$ to account for the activity-dependence of $A$ and $B$ (the value of $c$ is nearly 1, see SM Sec. S8 for further details). The lines in Fig.~\ref{reentranceVFT}(a) are the plots of Eq.~(\ref{reentrant}) with the values of $c$ as given in Table S1. The excellent agreement between the simulation results and Eq.~(\ref{reentrant}) demonstrates that it is indeed the critical surface shape that governs the re-entrant behavior.
\r{Indeed, neglecting the fit factor $c$ (see SM, Sec. S8), the key first two terms in the denominator of Eq. [10] just correspond to $\phi_c$. The competition between these terms determines, for given $f_0^2$ and $\phi$, the location $\tau_{p,\rm min}$ of the re-entrant minimum in $\tau_\alpha$ (see Fig. 3a). To understand qualitatively how $\tau_{p,\rm min}$ varies with $f_0$, notice from Fig. 3(b) that, for small $f_0$, $\phi_c$ is increasing with $\tau_p$ and hence $\tau_\alpha$ is decreasing, corresponding to $\tau_{p,\rm min}\to\infty$. For large $f_0$ the situation is reversed (Fig. 3d) and $\tau_{p,\rm min}\to 0$. In the intermediate range of $f_0$ where $\tau_{p,\rm min}$ is finite, it must therefore decrease with increasing $f_0$, consistent with the trend visible in Fig. 3(a).}

The re-entrant dynamics observed when $\tau_p$ increases is accompanied by a qualitative change in the physical relaxation process, since the system crosses over from near-equilibrium glassy relaxation dynamics when $\tau_p \to 0$ to non-thermal jamming physics when $\tau_p \to \infty$, as explained in Sec.~\ref{model}. This evolution from glass to jamming physics is captured by the behaviour of the self-intermediate scattering function shown in Fig.~\ref{reentranceVFT}(e), which we show here for constant $\phi=0.65$ and $f_0^2=0.045$ while varying $\tau_p$ over a broad range of five orders of magnitude. At small $\tau_p$, $F_s(k,t)$ shows the characteristic two-step relaxation decay typical of glassy relaxation. However, at very large $\tau_p$, $F_s(k,t)$ decays in a single step with no intermediate plateau, as usually observed in driven systems in the absence of thermal fluctuations \cite{ikeda2012,ikeda2013}. As expected, the MSD also shows a similar change of behavior with changing $\tau_p$ (SM Fig.~S6).
Beyond these qualitative changes, the non-monotonic dependence on $\tau_p$ of the final decay time of $F_s(k,t)$ is also evident from Fig.~\ref{reentranceVFT}(e).

Re-entrant behavior appears in a variety of systems, including colloidal suspensions of sticky hard spheres~\cite{Pham2002} and fluids confined within periodic potentials~\cite{Nandi2011, Mandal2014}. In these equilibrium colloidal systems, re-entrance is typically governed by large changes in the static structure. By contrast, re-entrant behavior in active systems has been described via changes in the caging dynamics~\cite{Liluashvili2017,Debets2021,Klongvessa2019} or via effective attractive interactions~\cite{Arora2022,Berthier2017}. Our work provides a complementary perspective on the mechanism underlying re-entrant behavior in active systems.
	
\section{Tunable fragility and sub-Arrhenius to super-Arrhenius crossover}

\label{fragility}

Several past studies have shown that activity may change the glass fragility of self-propelled systems~\cite{Mandal2016,Flenner2016,Nandi2018}. A tunable glass fragility was also reported in the context of vertex and Voronoi models of biological tissues \cite{sussman2018,li2021,sadhukhan2021,sadhukhan2024,sadhukhan2024AVM}. Here, we show that the glass fragility in our model depends on both $\tau_p$ and packing fraction $\phi$, with the physics being again controlled by the evolution of the glass critical surface constructed in Sec.~\ref{phasedia}.  

In thermal equilibrium, the glass fragility characterising slow dynamics is usually determined by following the temperature evolution of the structural relaxation time, $\tau_\alpha(T)$. Systems that exhibit a simple Arrhenius behavior are strong, whereas a stronger temperature dependence (also called super-Arrhenius) corresponds to fragile glasses. Fragility is graphically captured in Angell plots, where the logarithm of $\tau_\alpha$ is shown as a function of the inverse temperature so that data points for a strong glass lie on a straight line.  

To investigate fragility in active systems, we generalise this equilibrium analysis and follow the evolution of the structural relaxation time $\tau_\alpha$ as a function of the effective temperature $\Teff=f_0^2/(1+G\tau_p)$ defined in Eq.~(\ref{eq:Teff}). We tune $\Teff$ by varying $f_0^2$, at fixed values of $\phi$ and $\tau_p$. We construct the active analogue of Angell plots showing the logarithm of $\tau_\alpha$ as a function of inverse $\Teff$. We refine this representation by going to a rescaled version of the Angell plot, scaling $\Teff$ by its value $(\Teff)_g$ at the computer glass transition defined as 
$\tau_\alpha[(\Teff)_g]=10^3$. This scaling allows for a simpler visualisation of the evolution of the glass fragility with control parameters.

\begin{figure}
	\centering
	\includegraphics[width=8.6cm]{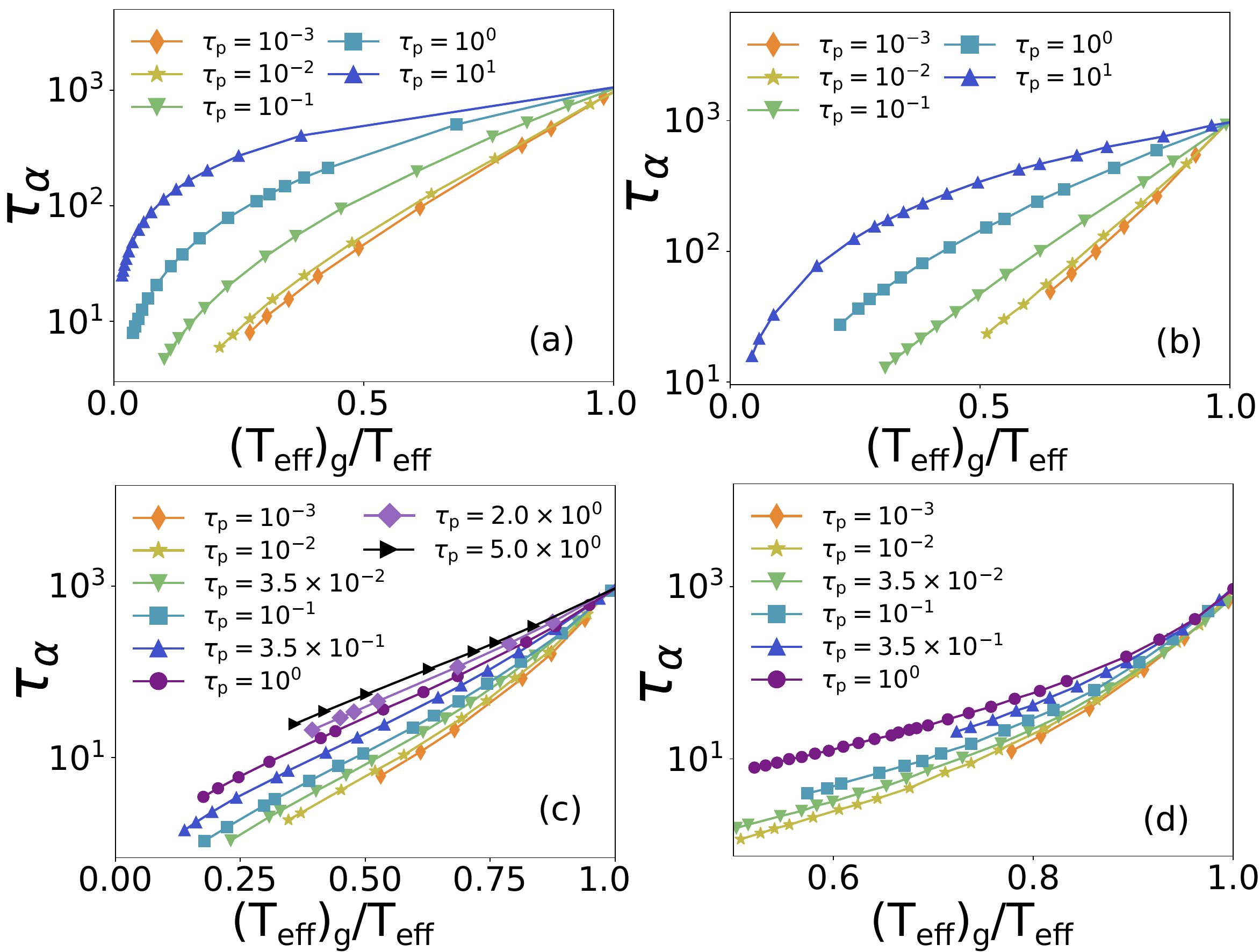}
	\caption{Evolution of the glass fragility shown using Angell plots. Each panel represents data obtained at a given packing fraction, $\phi$ with (a) $\phi = 0.625$, (b) $\phi= 0.650$, (c) $\phi= 0.693$, and (d) $\phi= 0.800$, and each panel contains data for a range of $\tau_p$ values. The glass fragility decreases systematically with increasing $\tau_p$, and increases systematically with increasing $\phi$. The behavior of $\tau_\alpha$ crosses over from sub-Arrhenius at low $\phi$ and/or large $\tau_p$ to super-Arrhenius for large $\phi$ and/or small $\tau_p$.}
	\label{sub_super}
\end{figure}

We collect the results of this analysis in the four panels of Fig.~\ref{sub_super}. Each panel represents an Angell plot constructed for a given packing fraction from $\phi=0.625$ to $\phi=0.800$, and the various curves in each panel are obtained for different values of the persistence time.  

These data reveal several intriguing features. In each panel, we observe that glass fragility always decreases when $\tau_p$ increases at a fixed density, a trend which holds at all densities. At the lowest density (Fig.~\ref{sub_super}a), we observe that all systems display sub-Arrhenius relaxation, that is, the relaxation time grows more slowly than in an Arrhenius fashion. This is not observed at large density (Fig.~\ref{sub_super}d), where all systems now display super-Arrhenius relaxation, very much like conventional passive molecular glass-forming materials. Therefore the glass fragility increases both when $\tau_p$ decreases and when $\phi$ increases, with a peculiar sub-Arrhenius regime found at low enough $\phi$ and large enough $\tau_p$.

A qualitatively similar evolution with the density of the glass fragility was observed previously in the Brownian limit, $\tau_p \to 0$, for a similar model of soft repulsive spheres~\cite{Berthier2009,Berthier2009-pre}. The microscopic explanation is relatively straightforward. When decreasing $T$ at constant $\phi$, the system ends up in an equilibrium hard sphere fluid if $\phi < \phi_d$. As a result, $\tau_\alpha (T \to 0)$ does not diverge, and this qualitatively explains the apparent sub-Arrhenius behaviour. Instead, above the critical density $\phi_d$, the system crosses a glass transition at a finite $T$, and as a consequence the relaxation dynamics appears to diverge at a finite $T$, which gives rise to a super-Arrhenius temperature dependence. In this view, the glass fragility is changing continuously with density, and it is primarily controlled by the distance to the critical density $\phi_d$ that characterises the $T \to 0$ hard sphere limit. Note that in this physical explanation of the evolution of fragility, the softness of the particles plays no role. In Brownian colloidal systems, particle softness has likewise been shown to play only a limited role in directly controlling glass fragility~\cite{philippe2018glass}. In some soft charged colloids, osmotic deswelling has been shown to produce a large fragility change~\cite{mattsson2009soft,pelaez2015impact}, but this is unrelated to our observations.

This qualitative interpretation easily extends to our observation in active systems, which we rationalise using the three-dimensional phase diagram in Fig.~\ref{threedphasedia}(b). When decreasing $\Teff$ (and thus $f_0^2$) at constant $\tau_p$ the system either ends in a fluid at $\phi$ below $\phi_d$, or in a glass at $\phi$ above $\phi_d$. This explanation is valid for any value of the persistence time, and it directly explains the $\phi$ dependence of the glass fragility reported in Fig.~\ref{sub_super}. In addition, since $\phi_d$ increases with $\tau_p$, the glass fragility observed at a given $\phi$ must decrease with $\tau_p$ because it is mostly controlled by the distance to $\phi_d$.

\section{Dynamic scaling near the hard sphere non-equilibrium glass transition}
\label{BerthierWitten}

\begin{figure*}
	\centering
	\includegraphics[width=17cm]{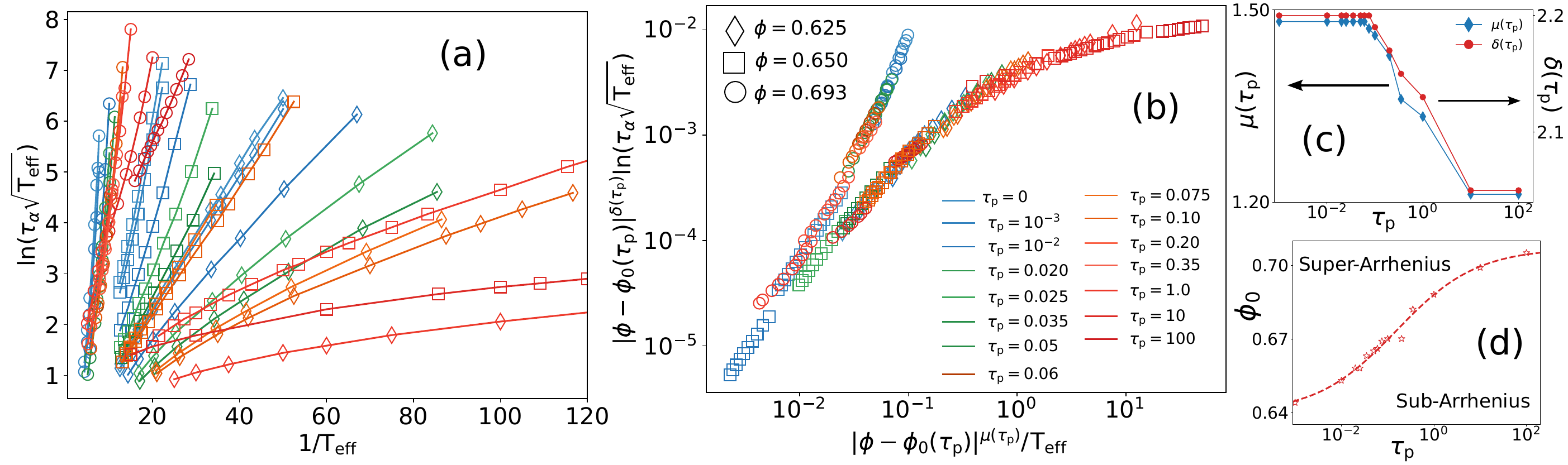}
	\caption{Dynamic scaling analysis collapses the glassy dynamics of active particles. (a) Angell plot using the rescaled relaxation time  $\tau_\alpha \sqrt{\Teff}$ as a function of $1/\Teff$. Different symbols are for different $\phi$, different colors are for different $\tau_p$.
          (b) Global data collapse along the two branches describing the $\phi<\phi_0$ sub-Arrhenius and $\phi > \phi_0$ super-Arrhenius family of curves, as described in Eq.~(\ref{eq:BW}).
          (c) The exponents $\delta$ and $\mu$ depart weakly from their equilibrium value as $\tau_p$ increases.
          (d) The critical hard sphere density $\phi_0(\tau_p)$ changes smoothly with $\tau_p$.}
	\label{plot_collapse}
\end{figure*}

For the equilibrium glassy dynamics of soft repulsive spheres~\cite{Berthier2009,Berthier2009-pre}, a dynamic scaling approach has previously been proposed to rationalize the qualitative variation across the $(\phi, T)$ plane. This analysis disentangles two aspects in the growth of $\tau_\alpha$. First, $\tau_\alpha$ grows at low $T$ simply because the thermal velocity of the system decreases, thus slowing down the relevant microscopic timescale $\tau_{\rm mic}$ controlling particle motion. For thermal systems, $\tau_{\rm mic} \propto 1/\sqrt{T}$, and it is thus convenient to rescale $\tau_\alpha$ by $\tau_{\rm mic}$ to single out the effect of glassiness.

The second, more interesting, cause for slow dynamics is the emergence of complex and cooperative glassy dynamics. For this part, the dynamic scaling amounts to first identifying the physical behaviour in the hard sphere limit ($T \to 0$), and to then assuming that thermalised soft spheres essentially obey the same physics as hard spheres, but at an ``effective'' value $\phi_{\rm eff}<\phi$ of the packing fraction, so that thermal soft spheres essentially appear as ``small'' hard spheres. Mathematically, the first assumption is a statement about the hard sphere $T \to 0$ limit, written as
\begin{equation}
  \tau_\alpha \sim \exp [ A / (\phi_0 - \phi)^\delta ] ,
  \label{eq:HS}
\end{equation}
which becomes equivalent to Eq.~(\ref{eq:VFT}) when $\delta = 1$. The connection between soft and hard particles then suggests the following scaling form:
\begin{equation}
\tau_\alpha \sim \exp \left[ \frac{A}{|\phi_0 - \phi|^\delta}  
F_{\pm} \left( \frac{ | \phi_0 - \phi |^\mu }{T}   \right)
\right]
\label{eq:BW}
\end{equation}
where two scaling functions $F_{\pm}(x)$ are introduced to describe the respective behaviour for densities above $\phi_0$, for $F_+(x)$, and below for $F_{-}(x)$. The hard sphere behaviour in Eq.~(\ref{eq:HS}) for $T \to 0$ imposes $F_{-}( x \to \infty) = 1$ and $F_{+}(x \to \infty) = + \infty$. Similarly, the continuity of the data for $\phi=\phi_0$ imposes that $F_{-}(x) \sim F_+(x) \sim x^{\delta / \mu}$ for $x\to 0$, so that $\tau_\alpha \sim \exp(A/T^{\delta/\mu})$ exactly at $\phi=\phi_0$.

The steps needed to extend the dynamic scaling analysis performed in equilibrium to active systems are relatively straightforward. The first step is to replace $T$ with the effective temperature $\Teff$ in Eq.~(\ref{eq:Teff}). In a second step, we rescale the relaxation time $\tau_\alpha$ with a microscopic timescale $\tau_{\rm mic} \sim 1 / \sqrt{\Teff}$. In a third step we generalise Eq.~(\ref{eq:BW}) by allowing the exponents $\delta$ and $\mu$, and the critical packing fraction $\phi_0$, to depend on the persistence time $\tau_p$. The scaling functions $F_{\pm}(x)$ could in principle also depend on $\tau_p$, but we find that this is not necessary to achieve a good collapse of the data. 

We now show how to apply this scaling procedure to our data. In Fig.~\ref{plot_collapse}(a), we show how the rescaled relaxation times $\tau_\alpha / \tau_{\rm mic} \sim \tau_\alpha \sqrt{\Teff}$ depend on the effective temperature $\Teff$ for multiple combinations of values of $\tau_p$ and $\phi$. In this rescaled form, the data at $\phi < \phi_0$ visibly saturate to a finite relaxation time in the $\Teff \to 0$ hard sphere limit, which leads to sub-Arrhenius temperature evolution. By contrast, the data for $\phi > \phi_0$ do not show any saturation and their evolution is compatible with a divergence at a finite effective temperature, which leads to super-Arrhenius temperature evolution.   

We are now in a position to apply Eq.~(\ref{eq:BW}) to our data. In practice, we find that the largest value of the packing fraction, $\phi=0.8$, is too far above the critical density $\phi_0$ and does not provide a good data collapse. Therefore, we did not use these data for the scaling analysis. The data collapse procedure is somewhat tedious as it requires the simultaneous identification of the free parameters $\delta$, $\mu$ and $\phi_0$ for each $\tau_p$. In practice, we initialized the fitting process with the values obtained in equilibrium~\cite{Berthier2009-pre} and slowly varied the fit parameters to achieve a satisfying result for all $\tau_p$ values. The outcome of this analysis is shown in Fig.~\ref{plot_collapse}(b), showing the data collapse of the relaxation data along the two branches described by $F_{-}$ and $F_{+}$, while the fitted parameters are shown in Figs.~\ref{plot_collapse}(c, d). The exponents $\delta$ and $\mu$ vary weakly with $\tau_p$ and show only small departures from their equilibrium values (corresponding to $\tau_p \to 0$). The critical density $\phi_0$ increases gradually with $\tau_p$ and saturates at large persistence times. As expected, it closely mirrors the evolution of $\phi_d$ discussed above (see SM Fig.~S4), as they both describe the same hard sphere dynamics using slightly different functional forms, Eqs.~(\ref{eq:VFT}, \ref{eq:HS}).       

The quality of the data collapse in Fig.~\ref{plot_collapse}(b) with weak variations of $(\delta, \mu)$ and scaling functions $F_{-}(x)$ and $F_{+}(x)$ that are independent of $\tau_p$ demonstrates that the scaling analysis proposed for equilibrium soft particles also applies to self-propelled soft particles. Only one physical quantity varies significantly with $\tau_p$ in this analysis, the critical density $\phi_0(\tau_p)$ that describes the $\Teff \to 0$ limit of self-propelled hard spheres. This promotes the ``glass point'' $\phi_0$ of Ref.~\cite{Berthier2009-pre} to a continuous ``glass line'' $\phi_0(\tau_p)$ with a dependence on the persistence time. A physical outcome of the data collapse in Fig.~\ref{plot_collapse} is the demonstration that glass fragility of active particles is directly controlled by the distance to the critical density $\phi_0(\tau_p)$, while the functional forms of the two scaling functions account for the crossover from sub-Arrhenius to super-Arrhenius.

\section{Discussion and perspectives}

\label{discussion}

We have studied the glassy dynamics in an active system of self-propelled soft spherical particles. This model system contains as limit cases thermal soft spheres and persistent self-propelled hard spheres, while for large $f_0$ and $\tau_p$ it describes self-propelled soft particles and so connects qualitatively also to the physics of confluent biological tissues. This broad range of physical behaviours captured by the model leads to a rich phenomenology, which we reveal here by performing a full exploration of the three-dimensional phase diagram $(f_0, \tau_p, \phi)$.

The construction and quantitative analysis of the three-dimensional phase diagram and its various limits allowed us to account for two non-trivial dynamic features: re-entrant glassy dynamics that emerges when $\tau_p$ is varied at fixed $\phi$ and $f_0$, and a glass fragility that is tuned by both changing $\tau_p$ and $\phi$. Our analysis generalises, and provides a simple interpretation for, related previous reports of anomalous dynamics in self-propelled particle systems~\cite{Berthier2017,Arora2022,Debets2021}. Our central conclusion is that the effect of activity is very well captured by the known equilibrium scaling description~\cite{Berthier2009-pre}, provided one promotes the hard sphere glass point $\phi_0$ to a hard sphere glass line $\phi_0(\tau_p)$ that depends continuously on the persistence time.

Interestingly, along this glass line of the active system one is effectively moving smoothly from a glass transition for small persistence times $\tau_p$ to a jamming transition for large $\tau_p$~\cite{Mandal2020,mandal2020aging,wiese2023,szamel2024} as shown by the disappearance of the intermediate plateau in the relaxation functions with increasing $\tau_p$. Microscopically, this behaviour can be rationalized by analogy with the physics of passive glasses subjected to mechanical deformation by steady shear~\cite{ikeda2012,ikeda2013}: here one also finds a smooth change between two distinct regimes, controlled by temperature. When thermal fluctuations are significant, corresponding to our active glasses at small $\tau_p$, relaxation under sufficiently slow shear proceeds by thermal activation across energy barriers. For times shorter than the barrier crossing time, particles can only relax partially by ``rattling'' in cages formed by their neighbours, causing plateaus in typical relaxation functions. In the athermal regime, on the other hand, relaxation is driven by barriers disappearing via saddle-node bifurcations as the energy landscape is gradually deformed by the applied shear. For our active systems, this is analogous to the slow tilting of the energy landscape by active forces~\cite{mandal2020aging,mandal2021,keta2023} in the large $\tau_p$ regime; either way, the relaxation has no analog of particles rattling in cages at early times, and relaxation functions therefore do not show plateaus.

An interesting perspective for future research is to connect the behaviour of soft active particles we have studied here to the physics revealed by studies of model systems for confluent epithelial tissues. In these models, the packing fraction $\phi$ does not control the physics as in soft particles as it is effectively fixed to unity. Instead there is a structural parameter controlling the behaviour of the system, the so-called target perimeter $p_0$~\cite{bi2016,sadhukhan2021,activereview,atia2021}. This drives the system from fluid-like states at large $p_0$ to solid-like behaviour at low $p_0$, in a manner analogous to $1/\phi$ in particle systems. In spite of these important differences, the parameters $p_0$ and $\phi$ play similar roles in so far as they control the transition from fluid to solid response, in the absence of thermal fluctuations and active forces. Intriguingly, there are several reports of sub-Arrhenius to super-Arrhenius crossover in the literature as $p_0$ is varied~\cite{sussman2018,sadhukhan2021,li2021,sadhukhan2024}, suggesting a possible analogy with soft particles. Re-entrant dynamics also exists in variants of these model systems~\cite{arora2024}. Future work should explore whether the analogy can be made more quantitative, and whether the analysis carried out here can also be useful to rationalise the characteristics of the glassy dynamics of biological tissues and its interplay with jamming physics, thus hopefully illuminating the role of many-body forces and confluence in the physics of tissue models.

\section*{acknowledgement}
PP and SKN acknowledge the support of the Department of Atomic Energy, Government of India, under Project identification No. RTI 4007. LB acknowledges the support of the French Agence Nationale de la Recherche (ANR), under grants ANR-20-CE30-0031 (project THEMA) and ANR-24-CE30-0442 (project GLASSGO). PS acknowledges funding by the Deutsche Forschungsgemeinschaft (DFG, German Research Foundation) under Project-ID 449750155, RTG 2756, Project A5. PS and SKN thank the Erwin Schr{\"{o}}dinger International Institute for Mathematics and Physics (ESI), University of Vienna (Austria) for the invitation to participate in the Thematic Program ``Linking Microscopic Processes to the Macroscopic Rheological Properties in Inert and Living Soft Materials'' in 2024 where part of this work has been carried out.


\onecolumngrid 


\section*{Supplementary material}

# Supplementary Material for "Scaling the glassy dynamics of active particles: Tunable fragility and reentrance"

Puneet Pareek,[1] Peter Sollich,[2] Saroj Kumar Nandi,[1] Ludovic Berthier[3]

*1. Tata Institute of Fundamental Research, Hyderabad - 500046, India*
*2. Institute for Theoretical Physics, University of Göttingen, Friedrich-Hund-Platz 1, 37077 Göttingen, Germany*
*3. Gulliver, CNRS UMR 7083, ESPCI Paris, PSL Research University, 75005 Paris, France*

## SI. Simulation details

We explore the effects of activity, with self-propulsion force $f_0$ and persistence time $\tau_p$, on a dense system of soft repulsive spheres. We use active Orstein-Uhlenbeck particles (AOUP) for activity. We take a three-dimensional binary system of 50:50 particle number ratio to avoid crystallization and the particles interact via the Weeks-Chandler-Andersen (WCA) potential, defined as

$$V_{\alpha\beta}(r) = 4\epsilon \left[ \left( \frac{\sigma_{\alpha\beta}}{r} \right)^{12} - \left( \frac{\sigma_{\alpha\beta}}{r} \right)^{6} \right] \tag{S1}$$

for $r \leq 2^{1/6}\sigma_{\alpha\beta}$ and zero otherwise. In Eq. (S1), $\alpha, \beta$ denote the particle species $A$ or $B$, $\epsilon = 1$ sets the unit of energy, $\sigma_{AA} = 1.4, \sigma_{AB} = 1.2$, and $\sigma_{BB} = 1.0$. We use molecular dynamics simulations to evolve the particle positions, $\mathbf{r}_i$, of the $i$th particle, using the over-damped equation of motion as

$$\dot{\mathbf{r}}_i = \xi_0^{-1} \left[ \mathbf{F}_i + \mathbf{f}_i \right], \tag{S2}$$

where $\mathbf{F}_i$ is the inter-atomic force on the $i$th particle, and $\mathbf{f}_i$ is the active force,

$$\tau_p \dot{\mathbf{f}}_i = -\mathbf{f}_i + \boldsymbol{\eta}_i, \tag{S3}$$

with $\langle \boldsymbol{\eta}_i(t)\boldsymbol{\eta}_j^{\mathrm{T}}(t') \rangle = 2\xi_0 f_0^2 \mathbf{I} \delta_{ij}\delta(t - t')$, $\mathbf{I}$ being the unit tensor and $\xi_0$, the friction coefficient, We set $\xi_0 = 1$ and present the results using the natural time unit, $\xi_0 \sigma_{BB}/\epsilon$.

We used LAMMPS [S1] for our simulations. The forward Euler scheme was used for numerically solving the AOUP equations. The timestep ($dt$) is kept at least 10 times less than the persistence time $\tau_p$. We chose a cubic box such that the packing fraction $\phi = 2^{1/2}\pi N(\sigma_{AA}^3 + \sigma_{BB}^3)/(12V)$ can be obtained by adjusting the volume $V$ of the system. We simulate the system long enough, typically ten times the relaxation time of the system, to reach steady state before starting the production run. We take into account the centre of mass correction while tracking the positions of the particles.

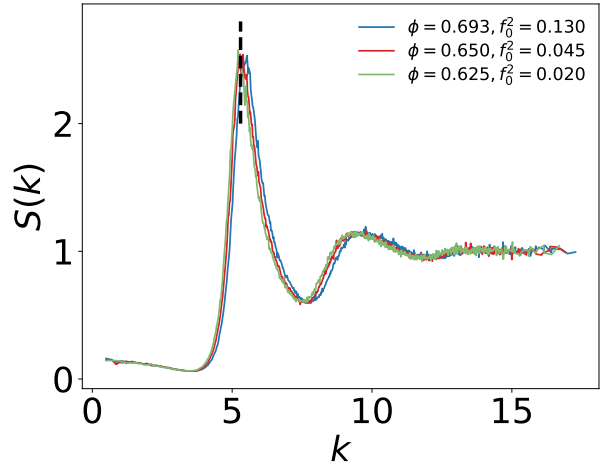

FIG. S1. The static structure factor at various $\phi$. The dashed line shows the peak position corresponding to $\phi = 0.65$.

## SII. Static structure factor of the passive system for various $\phi$

Figure S1 shows the static structure factor, $S(k)$, as a function of the wavevector $k$, for various $\phi$ and $f_0^2$ for the passive system at $\tau_p \to 0$. Note that the position of the first maximum does not change much for the density changes considered here. We have probed the dynamics at $k = 5.3$, corresponding to the peak position for $\phi = 0.65$.

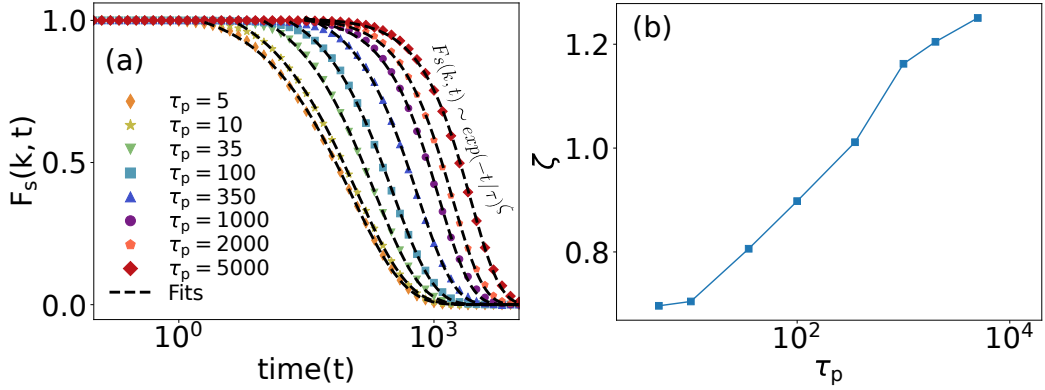

FIG. S2. (a) We fit the data of $F_s(k, t)$ with the form $\sim \exp[-(t/\tau)^\zeta]$ for different values of $\tau_p$ and $\phi = 0.625$. (b) The value of $\zeta$ becomes larger than unity in the large $\tau_p$ regime, implying a compressed exponential relaxation.

## SIII. Compressed Exponential

The stretching exponent, $\zeta$, for the decay of the self-intermediate scattering function $F_s(k, t)$ is defined via $F_s(k, t) \sim \exp[-(t/\tau)^\zeta]$, with $\zeta < 1$ giving a stretched exponential decay and $\zeta > 1$ a compressed exponential. We find that $\zeta$ has a $\tau_p$-dependence in active systems. Similar to thermal glasses, $\zeta < 1$ at small $\tau_p$. However, the value of this exponent increases with $\tau_p$. This growing value of $\zeta$ with increasing $\tau_p$ correlates with the shrinking and eventual disappearance of the sub-diffusive regime in the MSD for large $\tau_p$ [S2, S3].

## SIV. $a(\tau_p)$ and $\beta(\tau_p)$

We obtained the critical lines in the $(f_0^2 - \phi)$ plane at constant $\tau_p$ separating the liquid and the glass phases (Fig. 2c in the main text). We fit the simulation data for this phase diagram with

$$f_0^2 = a(\tau_p)[\phi_c - \phi_d(\tau_p)]^{\beta(\tau_p)/2} \tag{S4}$$

and obtain the parameters $a(\tau_p)$ and $\beta(\tau_p)$. We show the functional dependence of the parameters in Fig. S3.

## SV. Variation of $\phi_d$ and $\phi_0$ with $\tau_p$

In the main text we defined the location of the glass transition for the persistent hard sphere system as $\phi_d(\tau_p) = \phi_c(f_0^2 \to 0)$. We obtain $\phi_d(\tau_p)$ from the fit of Eq. (8) in the main text to the simulation data. Figure S4 shows the resulting behavior of $\phi_d(\tau_p)$. The line in Fig. S4 is a fit with the form $\phi_d(\tau_p) = a_d + b_d \tau_p^{0.67}/(1 + c_d \tau_p^{0.67})$, where $(a_d, b_d, c_d) = (0.595695, 0.232197, 3.26217)$.

As mentioned in the main text, the behavior of $\phi_0$ (Fig. 5d) is similar to that of $\phi_d$, the line in Fig. (5d) is a fit with the form $\phi_0(\tau_p) = a_0 + b_0 \tau_p^{0.5}/(1 + c_0 \tau_p^{0.5})$, where $(a_0, b_0, c_0) = (0.639654, 0.171824, 2.55624)$.

## SVI. Estimation of $T_{eff}$

We have introduced the effective temperature $T_{eff} = f_0^2/(1 + G\tau_p)$ in the main text. Reference [S4] showed that the effective temperature description agrees well with the long-time fluctuation-dissipation ratio, obtained via mode-

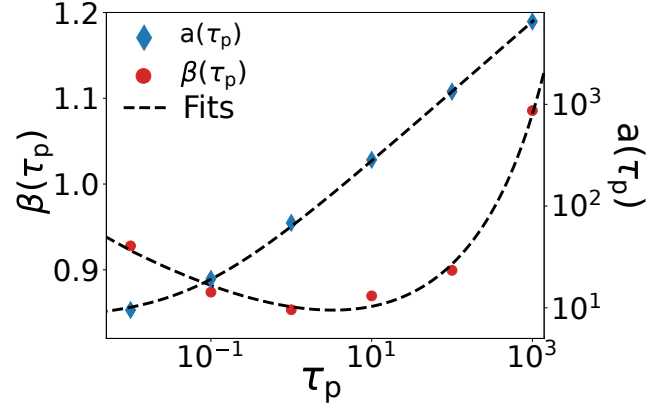

FIG. S3. The parameters $a(\tau_p)$ and $\beta(\tau_p)$ in Eq. (S4) with varying $\tau_p$. $a(\tau_p)$ increases at large $\tau_p$ whereas $\beta(\tau_p)$ shows a weakly non-monotonic behavior. The lines represent the following functions: $a(\tau_p) = 55.26\tau_p^{0.690} + 7.827$ and $\beta(\tau_p) = 2/(2.578\tau_p^{0.030} - 0.244\tau_p^{0.244})$.

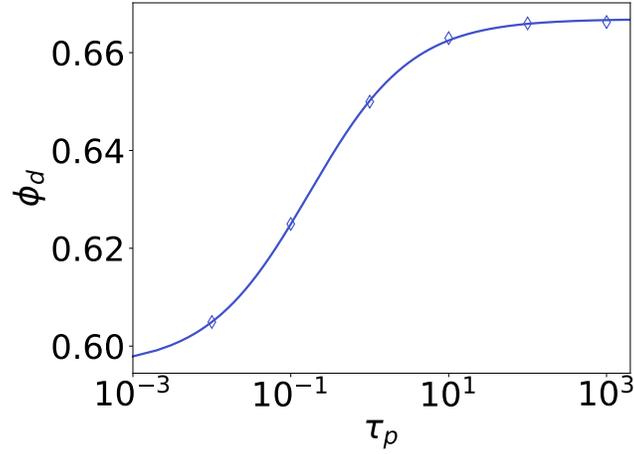

FIG. S4. $\phi_d$ shows a monotonic increase with $\tau_p$ and saturates in the two extreme limits of $\tau_p$.

coupling theory (MCT). Here, we used the MCT scaling form, $\tau_\alpha \sim \tau_0(T_{\text{eff}})^{-\gamma}$, where $\gamma$ is the MCT power law exponent for the equilibrium system, to fit the simulation data. We show these fits in Fig. S5; they give $G \approx 0.5$.

## SVII. Mean-square displacement (MSD) at various $\tau_p$

The mean-square displacement (MSD) changes its behavior with increasing $\tau_p$. When $\tau_p$ is small, the dynamics is dominated by the physics of glasses. MSD shows a characteristic sub-diffusive plateau at intermediate times before becoming diffusive at long times (Fig. S6). By contrast, when $\tau_p$ is large, the dynamics is dominated by the physics of jamming. MSD shows no sub-diffusive plateau at intermediate times, instead crossing over directly from a ballistic to a diffusive regime (Fig. S6).

## SVIII. Re-entrance relaxation time fits for $\phi = 0.625$ and $\phi = 0.693$

From the VFT fits, we obtain the liquid-glass critical lines as

$$\phi_c(f_0^2, \tau_p) = [f_0^2/a(\tau_p)]^{\beta(\tau_p)/2} + \phi_d(\tau_p). \tag{S5}$$

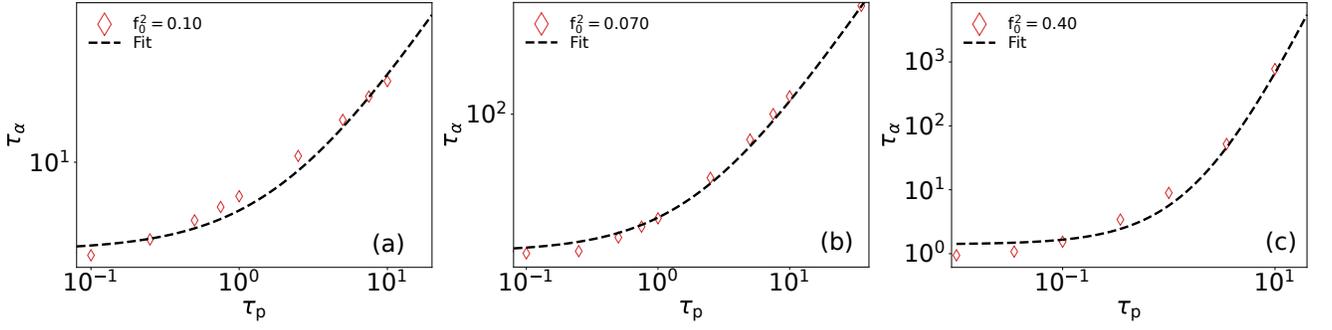

FIG. S5. Estimation of $G$ for $T_{\text{eff}}$. We fit the MCT scaling form, $\tau_\alpha = \tau_0(f_0^2/(1 + G\tau_p))^{-\gamma}$ with $\tau_0$, $\gamma$, and $G$ as fitting parameters. We find that $G = 0.5$ gives reasonable fits with $\tau_\alpha$ at various $\phi$, although $\gamma$ depends on $\phi$. The symbols show the simulation data and the dashed lines show the fits: (a) $\phi = 0.625$, $\gamma = 1.0$, (b) $\phi = 0.65$, $\gamma = 1.0$, (c) $\phi = 0.693$, $\gamma = 3.4$.

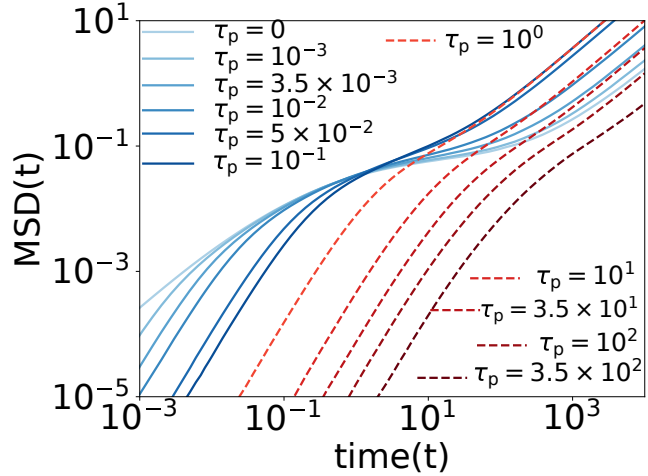

FIG. S6. The behavior of MSD at various $\tau_p$. MSD shows sub-diffusive plateau at small $\tau_p$, however, this plateau disappears as $\tau_p$ becomes large. We have used $\phi = 0.65$ and $f_0^2 = 0.045$ for these simulations.

Using this expression in the VFT formula, $\ln \tau_\alpha = A + B/(\phi_c(f_0^2, \tau_p) - \phi)$, we obtain

$$\ln \tau_\alpha = A(f_0^2, \tau_p) + \frac{B(f_0^2, \tau_p)}{[f_0^2/a(\tau_p)]^{\beta(\tau_p)/2} + \phi_d(\tau_p) - \phi}, \tag{S6}$$

where we have made the activity dependence of the parameters $A$ and $B$ explicit. Alternatively, we can include a parameter $c$ in the VFT equation to account for the variation in $A$ and $B$:

$$\ln \tau_\alpha = A + \frac{B}{c[f_0^2/a(\tau_p)]^{\beta(\tau_p)/2} + \phi_d(\tau_p) - \phi}. \tag{S7}$$

We show below that both forms, Eq. (S6) and Eq. (S7) yield similar fits.

We first show the comparison of the simulation data for $\tau_\alpha$ with Eq. (S7). The fits for $\phi = 0.65$ are displayed in the main text. In Fig. S7, we show the simulation data and the corresponding VFT lines for two other packing fractions, $\phi = 0.625$ and $\phi = 0.693$.

The values of $c$ obtained from the fits of the VFT expression, Eq. (S7), for three values of $\phi$ are given below.

We now show that we can set $c = 1$ by keeping the dependence of $A$ an $B$ on $f_0^2$ and $\tau_p$. To illustrate this, we show the comparison of the VFT form, Eq. (S6), with the simulation data for $\phi = 0.650$.

For a specific value of $\phi$ and $\tau_p$, the behaviors of $\tau_\alpha$ remains equilibrium-like with varying $f_0^2$. We therefore expect that $A$ and $B$ should remain independent of $f_0^2$. For different values of $\tau_p$, we first fit Eq. (S6) with the data of $\tau_\alpha$

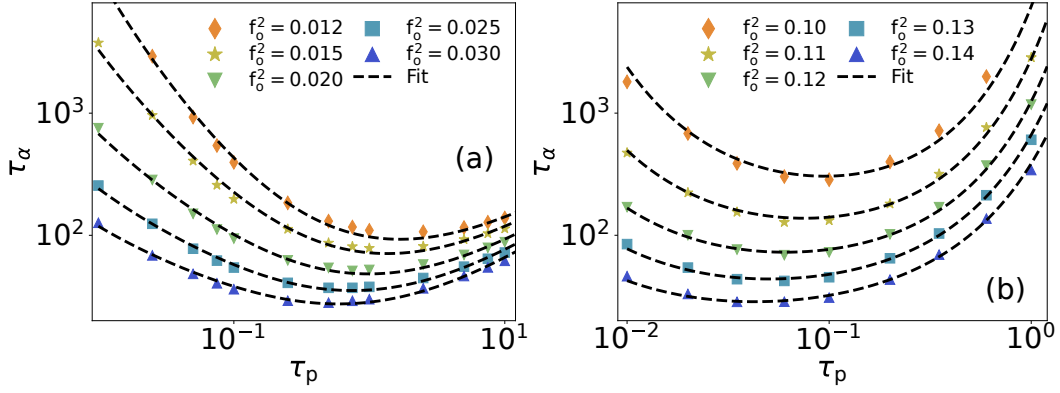

FIG. S7. We have plotted the fits according to Eq. (S7) for two other packing fractions, $\phi = 0.625$ and $\phi = 0.693$. The values for the parameters used are given in Table I.

TABLE I. Values of the parameters $A$, $B$, and $c$ corresponding to different values of $\phi$

| Value of $\phi = 0.625$, $A = -3.51354$, $B = 0.458038$. | | Value of $\phi = 0.650$, $A = -3.0504$, $B = 0.30175$. | | Value of $\phi = 0.693$, $A = -1.41854$, $B = 0.10146$. | |
|---|---|---|---|---|---|
| $f_0^2$ | $c$ | $f_0^2$ | $c$ | $f_0^2$ | $c$ |
| 0.012 | 1.2354 | 0.030 | 1.00769 | 0.10 | 0.833699 |
| 0.015 | 1.20079 | 0.035 | 0.984014 | 0.11 | 0.816152 |
| 0.020 | 1.16144 | 0.040 | 0.966882 | 0.12 | 0.80116 |
| 0.025 | 1.13154 | 0.045 | 0.952008 | 0.13 | 0.787542 |
| 0.030 | 1.10407 | 0.050 | 0.938523 | 0.14 | 0.775801 |

as a function of $f_0^2$ (Fig. S8a). From such fits for varying $\tau_p$, we obtain $A(\tau_p)$ and $B(\tau_p)$. We fit power laws to get analytical forms for $A(\tau_p)$ and $B(\tau_p)$ from these data [Figs. S8(b) and (c)]. Note that these analytical forms are used simply to approximate the trends of the parameters. Using these forms in Eq. (S6), we show the comparison of simulation data with Eq. (S6) as a function of $\tau_p$ in Fig. S8(d). A comparison of Fig. S8(d) with Fig. 3(a) in the main text shows that both forms, Eqs. (S6) and (S7), yield similar descriptions, although the latter is easier to fit to data.

---

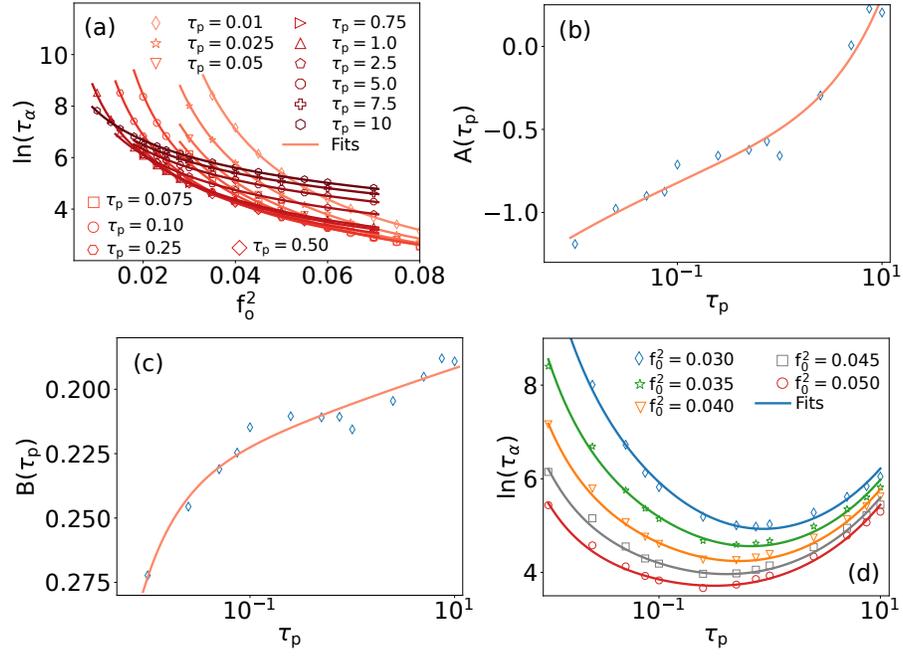

FIG. S8. (a) We fit Eq. S6 to the simulation data of $\tau_\alpha$ as a function of $f_0^2$ with $\phi = 0.65$ and different values of $\tau_p$. (b)-(c) Symbols represent the values of $A(\tau_p)$ and $B(\tau_p)$ from the fits. We obtain an analytical form by fitting power laws to the data; the lines represent $A(\tau_p) = 0.124\tau_p^{0.786} - 0.622\tau_p^{-0.132}$ and $B(\tau_p) = 0.0002\tau_p^{-1.181} + 0.205\tau_p^{-0.030}$. (d) Comparison of the VFT form, Eq. (S6) with the simulation data for $\tau_\alpha$ as a function of $\tau_p$. We have used the analytical forms for $A(\tau_p)$ and $B(\tau_p)$ and the lines represent Eq. (S6) without any fitting parameters, showing remarkable agreement with the simulation data (symbols).



\end{document}